\documentclass[pra,aps,twocolumn,showpacs,superscriptaddress,amsmath,amssymb,superscriptaddress]{revtex4}
\usepackage{graphicx}
\input{epsf}
\begin{document}
\epsfverbosetrue
\def\la{{\langle}}
\def\ra{{\rangle}}
\def\vep{{\varepsilon}}
\newcommand{\beq}{\begin{equation}}
\newcommand{\eeq}{\end{equation}}
\newcommand{\beqa}{\begin{eqnarray}}
\newcommand{\eeqa}{\end{eqnarray}}
\newcommand{\da}{^\dagger}
\newcommand{\wh}{\widehat}
\newcommand{\os}[1]{#1_{\hbox{\scriptsize{osc}}}}
\newcommand{\cn}[1]{#1_{\hbox{\scriptsize{con}}}}
\newcommand{\sy}[1]{#1_{\hbox{\scriptsize{sys}}}}

\title{Decoherence in a quantum harmonic oscillator monitored by a Bose-Einstein condensate}
\author{S. Brouard}
\email{sbrouard@ull.es}
\affiliation{Departamento de F\'{\i}sica Fundamental II, Universidad de La Laguna,
La Laguna E38204, Tenerife, Spain.}
\affiliation{Instituto Universitario de Estudios Avanzados (IUdEA).}
\author{D. Alonso}
\email{dalonso@ull.es}
\affiliation{Instituto Universitario de Estudios Avanzados (IUdEA).}
\affiliation{Departamento de F\'{\i}sica Fundamental y Experimental, Electr\'onica y Sistemas, 
Universidad de La Laguna, La Laguna E38204, Tenerife, Spain.}
\author{D. Sokolovski}
\affiliation{Department of Chemical Physics, University of the Basque Country, Leioa, Spain.}
\affiliation{IKERBASQUE, Basque Foundation for Science, 48011, Bilbao, Spain.}

\begin{abstract}
We investigate the dynamics of a quantum oscillator, whose evolution is monitored by a Bose-Einstein condensate (BEC) trapped in a symmetric double well potential. It is demonstrated that the oscillator may experience various degrees
of decoherence depending on the variable being measured and the state in which the BEC is prepared. These range from a
`coherent' regime in which only the variances of the oscillator position and momentum are affected by measurement,
to a slow (power law) or rapid (Gaussian) decoherence of the mean values themselves.
\end{abstract}
\date{\today}

\maketitle

\vskip0.5cm

In the past few years there has been much interest, both theoretical and experimental, in 
nano-mechanical oscillators whose quantum behaviour can be observed (measured) within the
limits imposed by the uncertainty relations \cite{schwab2005a,oconnell20101}. Typically,
various degrees of coherent control over such oscillators can be achieved by incorporating
them in hybrid devices involving superconducting microwave cavities \cite{regal2008}, 
superconducting qubits \cite{armour2008a,blencowe2008a}, single electron transistors
\cite{blencowe2005a,gurvitz2008a} and point contacts (PCs) \cite{PC}. More recently,
several schemes for coupling a quantum system to a Bose-Einstein condensate (BEC) have
been proposed \cite{BEC}. In the case of a measurement involving a PC in a large bias regime,
interaction between an oscillator and the electron current damps the latter leaving it in an
equilibrium thermal state \cite{mozyrsky2002a}. In this Letter we analyse a setup in which an
oscillator is coupled to a BEC trapped in a symmetric double well potential rather than to a
PC. With the atomic current dependent on the oscillator coordinate, the BEC is able to monitor
the oscillator evolution, at the cost of introducing decoherence to the oscillator dynamics.
This decoherence is the main subject of this Letter. We will show that an oscillator monitored
by a BEC  does not, in general, undergo a quantum to classical transition \cite{mozyrsky2002a,katz2007a}
and may, in some cases, retain a degree of coherence. For recent relevant work on the types of
decoherence possible in open systems we refer the reader to Ref. \cite{DECOH}.

We consider a system described by the Hamiltonian which is a generalisation of the `gatekeeper'
model introduced in \cite{sokolovski2009a} (we put $\hbar=1$)
\begin{figure}[ht]
\centering{\includegraphics[width=4cm]{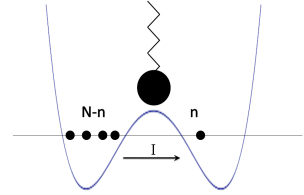}}
\caption{Double-well potential containing the $N$ bosons that may tunnel from one well to the other.
The flow of bosons is modulated by some function of the position of the oscillator.}
\label{system}
\end{figure}
\begin{equation}
H=\os H + \cn H +\delta\Omega\, \os A \otimes \cn B,
\end{equation}
where
\begin{eqnarray}
	\cn H&=&\Omega_0\,(c^{\dagger}_L c_L+c^{\dagger}_R c_R)
              -\Omega_1\,(c^{\dagger}_L c_R+c^{\dagger}_R c_L), \,\,\label{eq:1}\nonumber\\
	\cn B&=&c^{\dagger}_L c_R+c^{\dagger}_R c_L,
\end{eqnarray}
$c^{\dagger}_{L} (c^{\dagger}_{R})$ are the creation operators for the particles of the condensate
in the left (right) reservoir and $\os H$ is the oscillator Hamiltonian. The operator $\os A$
represents the variable which controls the tunnelling rate between the two wells, so that its
evolution can be monitored by observing the atomic current or a change in the number of bosons
in one of the reservoirs. Coupled in such a manner, the BEC shares with a conventional von Neumann
meter \cite{VN} the property that if prepared in a stationary state, it will remain in that state
inducing an additional force on the measured oscillator \cite{NONDEM}. With $\cn H$ and $\cn B$
commuting, $[\cn H,\cn B]=0$, such states are easily found to be 
\begin{equation}
|\tilde{\phi}_n \rangle=\frac{\left(c^{\dagger}_L+c^{\dagger}_R\right)^{N-n}\,
\left(c^{\dagger}_L-c^{\dagger}_R\right)^n}{\sqrt{2^N (N-n)!\, n!}}\left|0\cn{\right>},\,(n=0,...,N),
\end{equation}
where the vacuum $\left|0\cn{\right>}$ corresponds to no bosons in the condensate. Assuming that the
oscillator and the BEC are prepared in a product state, $\rho(0)=\os \sigma(0) \otimes \cn \rho(0)$,
and noting that $\cn B |\tilde{\phi}_n \rangle=(N-2 n)|\tilde{\phi}_n \rangle$, we find the state of
the monitored oscillator at a time $t$ by tracing out the BEC degrees of freedom,
%
\begin{equation}
\os\sigma(t)=\sum_n P_n \os\sigma^{(n)}(t), \quad  P_n=\cn{\rm Tr}[|\tilde{\phi}_n\rangle \langle
\tilde{\phi}_n| \cn \rho(0)],
	\label{eq:5}
\end{equation}
where
\begin{equation}\label{eq:4}
\os\sigma^{(n)}(t)\equiv e^{-i\os{\mathcal{H}} (\epsilon_n)t} 
\os \sigma(0) e^{i\os{\mathcal{H}} (\epsilon_n)t},
\end{equation}
$\os{\mathcal{H}}(\epsilon)\equiv\os H+\epsilon\os A$, and $\epsilon_n\equiv \delta\Omega(N-2n)$.

Thus, $\os\sigma(t)$ is an incoherent superposition of the states obtained by evolving $\os \sigma(0)$
with the family of Hamiltonians $\os{\mathcal{H}}(\epsilon_n)$, $n=0,1,...,N$, weighted by the probabilities
$P_n$ to find the BEC in the state $|\tilde{\phi}_n\rangle$. Accordingly, at a time $t$, the expectation value
of an oscillator variable represented by an operator $\os O$ is given by the sum
\begin{equation}
\langle \os O \rangle= \sum_n P_n {\os {\rm Tr}}\left[\os\sigma^{(n)}(t) \os O \right].
	\label{eq:6}
\end{equation}
Following \cite {sokolovski2009a}  we take the limit in which the number of atoms becomes large, while the
coupling between the oscilllator and each individual atom is reduced, namely
\begin{equation}\label{irrev}
N\to\infty, \quad \delta\Omega\to 0, \quad \delta\Omega\sqrt{N}=\kappa,
\end{equation}
and choose $\Omega_1=0$ so as to exclude a constant background current.
The conditions (\ref{irrev}) ensure that a macroscopic atomic current flows from the left to the right well,
while the Rabi period of an individual atom tends to infinity. Thus, the atoms are not going to return to their
initial state in the foreseeable future, i.e., the BEC becomes an irreversible meter with a large number of
degrees of freedom \cite{sokolovski2009a,sokolovski2009b}. Replacing sums by integrals, $2\delta\Omega \sum_n
\rightarrow \int_{-\infty}^{\infty}d\epsilon$, we rewrite (\ref{eq:5}) as  
\begin{eqnarray}\label{sigma}
\os \sigma(t)&=&
\int_{-\infty}^{\infty} d\epsilon\,P(\epsilon)
\left\{
\sum_i \left<\psi^{\epsilon}_i|\os \sigma(0)|\psi^{\epsilon}_i\right>
\left|\psi^{\epsilon}_i\right>\left<\psi^{\epsilon}_i\right|\right.\nonumber\\
&+&\left.\sum_{i\ne j} e^{-i\left(E^{\epsilon}_i-E^{\epsilon}_j\right)t}
\left<\psi^{\epsilon}_i|\os \sigma(0)|\psi^{\epsilon}_j\right>
\left|\psi^{\epsilon}_i\right>\left<\psi^{\epsilon}_j\right|\right\},
\end{eqnarray}
where $E^{\epsilon}_i$ and $\left|\psi^{\epsilon}_i\right>$ are the eigenvalues and eigenvectors respectively
of the Hamiltonian $\os{\mathcal{H}}(\epsilon)$ and $P(\epsilon_n)\equiv P_n/(2\delta\Omega)$. The long time
behaviour of $\os \sigma(t)$ now depends on the spectra $E^{\epsilon}_i$. Indeed, for $E^{\epsilon}_i-
E_j^{\epsilon}\ne const(\epsilon)$ rapidly oscillating exponentials will cause the second term in Eq.(\ref{sigma})
to vanish, so that $\os \sigma(t)$ [and with it the averages (\ref{eq:6})] will tend to stationary values as
$t\rightarrow \infty$. Without such a cancellation, the oscillator will not be able to reach a steady state
no matter how long one waits. 

Consider further a BEC initially localised in the left well, $\cn\rho(0)=|\phi_0\rangle\langle\phi_0|$, where
$|\phi_0 \rangle=(c^{\dagger}_L)^N\cn{|0\rangle}/\sqrt{N!}$, coupled to a harmonic oscillator of mass $m$ and
frequency $\omega_0$, $\os H=\omega_0\left(a^{\dagger}a+\frac{1}{2}\right)\equiv P^2/2m+\frac{1}{2}m\omega_0^2
X^2$, charged so as to affect the barrier between the two wells and the tunnelling rate. Both couplings linear
($\os A \sim X$) and quadratic ($\os A \sim X^2$) in the oscillator coordinate are possible \cite{tian2004a}.
For the probability weights $P_n$ in Eqs. (\ref{eq:5}) and (\ref{eq:6}) we have 
\begin{equation}\label{PN}
P_n={N!}/(2^N (N-n)! n!),
\end{equation}
and applying the Stirling formula in the limit (\ref{irrev}) yields $P(\epsilon)=(2\pi\kappa^2)^{-1/2}
\exp(-\epsilon^2/2\kappa^2)$. Rather than analyze the density matrix (\ref{eq:5}) it is convenient to consider
the mean position and momentum of the monitored oscillator, together with their variances, thus choosing the
operator $\os O$ in Eq.(\ref{eq:6}) to be $X$, $X^2$, $P$ or $P^2$.

\begin{figure}[ht]
\centerline{\includegraphics[width=7.5cm, angle=-90]{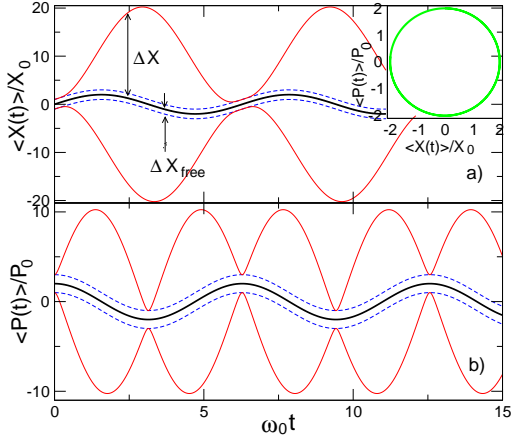}}
\caption{Coherent motion ($X$ is monitored) of a coherent  initial oscillator state with 
$\left<X(0)\right>/X_0=0$, $\left<P(0)\right>/P_0=2$;
$X_0\equiv (2m\omega_0)^{-1/2}$, $P_0\equiv (m\omega_0/2)^{1/2}$.
a) Mean position $\left< X(t)\right>/X_0$ (thick solid) vs. $\omega_0 t$.
Also shown are $[\left< X(t)\right>\pm \Delta X] /X_0$ (solid) and 
$[\left<X(t)\right>\pm \Delta X_{free}] /X_0$ (dashed);
b) Mean momentum $\left< P(t)\right>/P_0$ (thick solid) vs. $\omega_0 t$.
Also shown are $[\left< P(t)\right>\pm \Delta P] /P_0$ (solid) and 
$[\left<P(t)\right>\pm \Delta P_{free}] /P_0$ (dashed).
Inset: $\left< P(t)\right>/P_0$ vs. $\left< X(t)\right>/X_0$.}
\label{fig:2}
\end{figure}

{\it A. Coherent motion with `breathing'.} With $\os A\equiv a\da +a=\sqrt{2m\omega_0}X\equiv X/X_0$ the BEC
monitors oscillator's position $X$, and we need to consider motion in a family of harmonic potentials shifted
relative to the original one by $\delta x_n\equiv \sqrt{\frac{2} {m\omega_0}}\,\frac{\epsilon_n}{\omega_0}$,
$n=0,1,...,N$,
\begin{equation} \label{coh}
V_n(x) = m\omega_0^2(x+\delta x_n)^2/2 -m\omega_0^2\delta x_n^2/2.
\end{equation}
Thus, the energy differences in Eq.(\ref{sigma}) are independent of $\epsilon$, $E^{\epsilon}_i-E^{\epsilon}_j=
\omega_0 (i-j)$, and no steady state can be reached. We note further that since $\os{\mathcal{H}}(\epsilon)$ in
Eq.(\ref{eq:4}) remains quadratic in both $X$ and $P$, equations of motion for the five operators $X$, $P$,
$X^2$, $P^2$ and $XP+PX$ form a closed system which can be solved for each value $\epsilon_n$. Averaging the
results with the probabilities $P_n$ (c.f. Eq.(\ref{eq:6})) then yields 
\begin{eqnarray}
\frac{d\left<X(t)\right>}{dt}&=&\frac{1}{m}\left<P(t)\right>\\
\frac{d\left<P(t)\right>}{dt}&=&-m\omega_0^2²\left<X(t)\right>
-\sqrt{2m\omega_0}\sum_n P_n \epsilon_n.\nonumber
\end{eqnarray}
Probability distribution in Eq.(\ref{PN}) is symmetric about $N/2$ so that the sum in the last equation vanishes,
and we find the mean values of both the coordinate and the momentum unchanged by the presence of the BEC, $\langle
X(t)\rangle=\langle X(t)\rangle_{free}$ and $\langle P(t)\rangle=\langle P(t)\rangle_{free}$, where the subscript
`free' refers to an oscillator uncoupled from the BEC. This does not, however, imply that the BEC has no effect on
the dynamics of the oscillator. Indeed, calculating the variances we find
\begin{eqnarray}
	(\Delta X)^2 &=&\langle X^2\rangle-\langle X \rangle^2=(\Delta X)^2_{free}
+4\sigma_X^2\sin^4\left(\omega_0 t/2\right)\nonumber\\
	(\Delta P)^2 &=& \langle P^2\rangle-\langle P \rangle^2=(\Delta P)^2_{free}
+\sigma_P^2\sin^2(\omega_0 t),\nonumber
	\label{eq:8b}
\end{eqnarray}
where $\sigma_X\equiv\sqrt{\frac{2}{m\omega_0}}\,\frac{\delta\Omega\sqrt{N}}{\omega_0}$ and $\sigma_P\equiv
\sqrt{2m\omega_0}\,\frac{\delta\Omega\sqrt{N}}{\omega_0}$. Thus, while $\left<X(t)\right>$ and $\left<P(t)\right>$
follow their unperturbed trajectories, the widths of the corresponding distributions `breath', first increasing and
then decreasing again. Figure \ref{fig:2} shows the dynamics of the corresponding mean values and variances
for an oscillator prepared in a coherent state (minimal Gaussian wavepacket) $\langle x|\os \psi(0)\rangle=(m
\omega_0/\pi)^{1/4}e^{-m\omega_0[x-\langle X(0)\rangle]^2/2}e^{i \langle P(0)\rangle x}$ with $\delta \Omega^2
N/\omega_0^2=25$.
Note that $\Delta X(t)$ recovers its original value after every period $T=2\pi/\omega_0$ as the oscillator
returns to its initial state in each $V_n(x)$. The momentum variance $\Delta P(t)$ does so also after every
half-period, when the shape of the original wavepacket is restored but the position of its centre is reflected
with respect to the origin of each $V_n$, i.e., when $\Delta X(t)$ reaches its maximum value.

{\it B. Gaussian decoherence.} With $\os A\equiv (a\da +a)^2= X^2/X_0^2$ the BEC monitors the square of the
oscillator's position, $X^2$, and we need to consider motion in a family of harmonic potentials with the same
origin, but with different frequencies, $\omega_n=\sqrt{\omega^2_0+4\epsilon_n\omega_0}$, $n=0,1,...,N$,
\begin{equation}\label{incoh}
V_n(x) = m(\omega_0^2+ 4\epsilon_n\omega_0)x^2/2=m\omega_n^2X^2/2.
\end{equation}
Now $E^{\epsilon}_i-E^{\epsilon}_j=\left(i-j\right)\sqrt{\omega^2_0+4\epsilon\omega_0}\ne const(\epsilon)$, 
and we expect that in the irreversible limit (\ref{irrev}) oscillator will undergo a 
relaxation to the steady state given by the first term in Eq.(\ref{sigma}).
Solving equations of motion for each oscillator frequency $\omega_n$ and averaging with the probabilities 
(\ref{PN}) we find
\begin{eqnarray}\label{eq:11}
&&\langle X(t) \rangle=\sum_n P_n \left\{\langle X(0)\rangle\cos(\omega_n t)+
\frac{\langle P(0) \rangle}{m\omega_n}\sin(\omega_n t)\right\}\nonumber\\
&&\langle X^2(t) \rangle=\sum_n P_n \left\{
\frac{\langle (XP+PX)(0) \rangle}{2 m \omega_n}\sin\left(2 \omega_n t\right)\right.\\
&&\left.+\frac{\langle P^2(0)\rangle}{2 m^2 \omega_n^2}\left(1-\cos\left(2\omega_n t\right)\right)
+\frac{\langle X^2(0) \rangle}{2}\left(1 + \cos\left(2 \omega_n t\right)\right)\right\}\nonumber\\
&&\frac{\langle P(t) \rangle}{m\omega_n}=
\sum_n P_n \left\{\frac{\langle P(0) \rangle}{m\omega_n}  \cos(\omega_n t)-
\langle X(0) \rangle\sin(\omega_n t)\right\}\nonumber\\
&&\frac{\langle P^2(t) \rangle}{m^2\omega_0^2}=\sum_n P_n \left\{
-\frac{\langle (XP+PX)(0) \rangle}{2m\omega_n}\sin\left(2 \omega_n t\right)\right.\nonumber\\
&&\left.+\frac{\langle P^2(0) \rangle}{2m^2\omega_n^2}\left(1+\cos\left(2\omega_n t\right)\right)
+\frac{\langle X^2(0) \rangle}{2}\left(1 - \cos\left(2 \omega_n t\right)\right)\right\},\nonumber
\end{eqnarray}
where $\langle \os O(0) \rangle_= \os{\rm Tr}[\os{\it O}\os\sigma(0)]$ is the expectation value of $\os O$ in the
initial oscillator state. We note that for $\epsilon_n < -\omega_0/4$,  $\omega_n$ becomes imaginary as the
interaction turns the oscillator potential into a parabolic repeller, leading to the break up of the system.
Choosing the interaction to be small enough to neglect the possibility of breakup and taking the limit
(\ref{irrev}), we replace the sums in Eqs.(\ref{eq:11}) by integrals to obtain
\begin{eqnarray}\label{XP}
\langle X(t)\rangle&=&\langle X(0)\rangle\,{\rm Re}[f_0(\omega_0 t)]+\frac{\langle P(0)\rangle}{m\omega_0}\,
{\rm Im}[f_1(\omega_0 t)],\\
\langle P(t)\rangle&=&\langle P(0)\rangle\,{\rm Re}[f_0(\omega_0 t)]-m\omega_0\langle X(0)\rangle\,{\rm Im}
[f_{-1}(\omega_0 t)]\nonumber
\end{eqnarray}
with
\begin{equation}\label{FBETA}
f_{\beta=0,\pm 1}(\tau;\sigma)\equiv
\frac{1}{\sqrt{2 \pi \sigma^2}}\int_{-1}^{\infty}dz\, e^{-z^2/2\sigma^2}\,\,
\frac{e^{i \tau \sqrt{1+z}}}{(1+z)^{\beta/2}},
\end{equation}
where $\sigma=4\kappa/\omega_0$ and $\tau=\omega_0t$.
For a weak coupling, $\sigma<<1$, and times not exceeding 
$1/\omega_0 \sigma^2$,  the exponent in (\ref{FBETA}) can be expanded 
up to the second order in $z$. Replacing the lower limit of integration by
$-\infty$ and evaluating Gaussian integrals yields
\begin{equation}\label{FBETAa}
f_{\beta}(\tau;\sigma) =(1+i\tau\sigma^2/4)^{-1/2}e^{-\tau^2\sigma^2/(8+2i\sigma^2\tau)}
e^{i\tau}+O(\sigma). 
\end{equation}
It is readily seen that, irrespective of the value of $\beta$,  $|f_{\beta}(\omega_0 t)|$ decays 
on a time scale $T\equiv 1/(\omega_0\sigma) \approx \kappa^{-1}$ so that for
$T << t \simeq\omega_0/\kappa^2$ both the mean position $\langle X(t)\rangle$ and the mean momentum
$\langle P(t)\rangle$ will have decayed to zero, regardless of their initial values,
the decay being Gaussian in time.
\begin{figure}[t]
\centerline{\includegraphics[width=7.5cm, angle=-90]{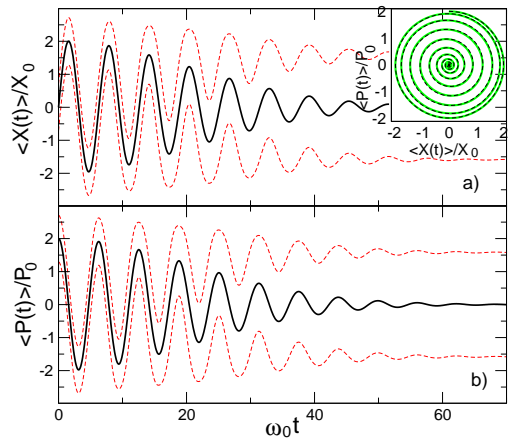}}
\caption{Gaussian decoherence ($X^2$ is monitored) of a coherent  initial oscillator state with 
$\left<X(0)\right>/X_0=0$, $\left<P(0)\right>/P_0=2$;
$X_0\equiv (2m\omega_0)^{-1/2}$, $P_0\equiv (m\omega_0/2)^{1/2}$.
a) Mean position $\left< X(t)\right>/X_0$ (thick solid) vs. $\omega_0 t$.
Also shown are $[\left< X(t)\right>\pm \Delta X] /X_0$ (dashed);
b) Mean momentum $\left< P(t)\right>/P_0$ (thick solid) vs. $\omega_0 t$.
Also shown are $[\left< P(t)\right>\pm \Delta P] /P_0$ (dashed).
Inset: $\left< P(t)\right>/P_0$ vs. $\left< X(t)\right>/X_0$
given by Eqs.(\ref{eq:11}) (thick solid) and (\ref{XP}-\ref{FBETAa}) (dashed).}
\label{fig:3}
\end{figure}
Figure \ref{fig:3} illustrates Gaussian decoherence for the same initial coherent state as in Fig. \ref{fig:2} 
and for $N=100$ and $\delta \Omega/\omega_0=0.0024$.
We note [and this is a general effect of a weak coupling, $\kappa<< \omega_0$, acting 
over a long time $t>>T$ , c.f. Eq.(\ref{eq:11})] that in the final steady state of the oscillator the initial
(kinetic) energy is shared equally between its kinetic and potential components, i.e., for $t>>T$ we have
 $\frac{\langle P^2(t)\rangle}{2m} = \frac{1}{2}m\omega_0^2\langle X^2(t)\rangle
 =\frac{\langle P^2(t=0)\rangle}{4m}$.

{\it C. Power law decoherence.} Other types of decoherence are possible with different choices 
of the initial state of the BEC. For example, a power law decoherence
can be achieved by replacing in Eq.(\ref{FBETA}) the smooth Gaussian factor by a discontinuous one. Thus, choosing
 $P_n\equiv|\langle\tilde{\phi}_n|\phi_0\rangle|^2=e^{-\alpha\omega_n}/\sum_{n\ge N/2}^Ne^{-\alpha\omega_n}$ for
$n\ge N/2$ and zero otherwise and taking the limit (\ref{irrev}) for a state with $ \langle X(0)\rangle =0$ yields
 \begin{eqnarray}\label {X}
\langle X(t)\rangle= 
 \langle P(0)\rangle \frac{\alpha\sin(\omega_0 t)+t
 \cos(\omega_0t)}{m(\alpha\omega_0+1)(1+t^2/\alpha^2)},
\end{eqnarray}
so that for $t>>\alpha$, $\langle X(t)\rangle$ tends to zero as $1/t^{}$. 

In summary, in a hybrid setup involving an oscillator and a BEC in a symmetric double well potential we have an example
of a quantum detector with a large number of degrees of freedom. In the irreversible limit the meter provides
unidirectional macroscopic atomic current whose magnitude depends on the oscillator's position. Unlike in the
case of a point contact, the measurement  does not lead to universal damping of the oscillator and eventual
thermalisation of its initial state. Rather,  depending on the oscillator variable being monitored as well as
on the initial state of  the BEC, the oscillator may o may not undergo relaxation to a steady state and
retain a degree of initial coherence. The absence of a quantum to classical transition predicted, for example,
for an oscillator coupled to a PC, is a consequence of the fact that a single energy level, rather than a broad
energy band, is available for each tunnelling boson.
\vspace*{0.2cm}

\begin{acknowledgments}
We are grateful to Shmuel Gurvitz for useful discussions. Two of us (S.B. and D.A.) acknowledge
financial support provided by Ministerio de Educaci\'on y Ciencia, Spain, (grants FIS2007-64018 and FIS2010-19998).
\end{acknowledgments}



\begin{thebibliography}{17}
\bibitem[{\citenamefont{Tian and Zoller}(2004)}]{NONDEM}
{Note that if the oscillator were to be
taken as the meter and the BEC as the measured system, a \emph{back-action evasion} condition is satisfied,
characteristic of a Quantum Non-Demolition (QND) measurement (see \cite{Braginsky1996a} and \cite{breuer2002a} for a review)}.
\end{thebibliography}


\end{document}